\documentclass[conference]{IEEEtran}
\usepackage{mathpazo}
\usepackage{times}
\usepackage{color}
\usepackage{amsmath}
\usepackage{amsfonts}
\usepackage{latexsym}
\usepackage{amssymb}
\usepackage{cite}

\usepackage{upref}
\usepackage{theorem}
\usepackage{graphicx}
\usepackage{psfrag}

\usepackage{hyperref}
\usepackage{mathrsfs}

\theoremstyle{plain}
\theorembodyfont{\normalfont\slshape}

\newtheorem{thm}{Theorem$\!$}
\newenvironment{theorem}
{\begin{thm}\hspace*{-1ex}{\bf.}}{\end{thm}}

\newtheorem{lem}[thm]{Lemma$\!$}
\newenvironment{lemma}{\begin{lem}\hspace*{-1ex}{\bf.}}{\end{lem}}

\newtheorem{prop}[thm]{Proposition$\!$}

\newtheorem{cor}[thm]{Corollary$\!$}

\newtheorem{defn}[thm]{Definition$\!$}

\newtheorem{xmpl}[thm]{Example$\!$}
\newenvironment{example}{\begin{xmpl}\hspace*{-1ex}{\bf.}}{\hfill$\Box$\end{xmpl}}

\newtheorem{cnstr}{Construction$\!$}
\newenvironment{construction}{\begin{cnstr}\hspace*{-1ex}{\bf.}}{\end{cnstr}}

\newcounter{enumrom}
\renewcommand{\theenumrom}{(\roman{enumrom})}

\makeatletter
\renewcommand{\@endtheorem}{\endtrivlist}
\makeatother

\makeatletter
\renewcommand{\thefigure}{{\@arabic\c@figure}}
\renewcommand{\fnum@figure}{{\bf Figure\,\thefigure}}
\makeatother

\newcommand{\mathset}[1]{\left\{#1\right\}}
\newcommand{\abs}[1]{\left|#1\right|}
\newcommand{\ceilenv}[1]{\left\lceil #1 \right\rceil}
\newcommand{\floorenv}[1]{\left\lfloor #1 \right\rfloor}
\newcommand{\parenv}[1]{\left( #1 \right)}

\newcommand{\be}[1]{\begin{equation}\label{#1}}
\newcommand{\ee}{\end{equation}}

\renewcommand{\le}{\leqslant}
\renewcommand{\leq}{\leqslant}
\renewcommand{\ge}{\geqslant}
\renewcommand{\geq}{\geqslant}

\renewcommand{\Bbb}{\mathbb}

\newcommand{\Cref}[1]{Co\-ro\-lla\-ry\,\ref{#1}}

\renewcommand{\Bbb}{\mathbb}

\newcommand{\N}{{\Bbb N}}
\newcommand{\Q}{{\Bbb Q}}

\DeclareMathOperator{\id}{Id}
\DeclareMathOperator{\per}{per}

\outer\def\proclaim #1. #2\par{\medbreak
 \noindent{\bf#1.\enspace}{\sl#2\par}%
 \ifdim\lastskip<\medskipamount \removelastskip\penalty55\medskip\fi}

\global\long\def\specialfont#1{\mathscr{\mathsf{#1}}}
\global\long\def\acm{\bar{\mathfrak{\specialfont M}}}
\global\long\def\wcm{\hat{\mathfrak{\specialfont M}}}
\global\long\def\acr{\bar{\mathfrak{\specialfont R}}}
\global\long\def\wcr{\hat{\mathfrak{\specialfont R}}}
\global\long\def\aca{\bar{\mathfrak{\specialfont A}}}
\global\long\def\wca{\hat{\mathfrak{\specialfont A}}}
\global\long\def\dk{\mathsf{d}_{\mathsf{K}}}
\global\long\def\dc{\mathsf{d}_{\mathsf{C}}}
\global\long\def\dist{\mathsf{d}}
\global\long\def\bk{\mathsf{B}_{\mathsf{K}}}
\global\long\def\bc{\mathsf{B}_{\mathsf{C}}}
\global\long\def\bb{\mathsf{B}}
\global\long\def\Sn{\mathbb{S}_n }
\global\long\def\Sm{\mathbb{S}_m }
\global\long\def\expl#1#2{\stackrel{\mathsf{(#1)}}{#2}}
\newcommand{\ppi}{\pi}

\begin{document}


\title{\Huge\bf Rate-Distortion for Ranking with \\Incomplete Information}


\author{
\IEEEauthorblockN{\textbf{Farzad Farnoud (Hassanzadeh)}}
\IEEEauthorblockA{Electrical Engineering  \\
California Institute of Technology \\
Pasadena, CA 91125, U.S.A. \\
{\it farnoud@caltech.edu}}
\and
\IEEEauthorblockN{\textbf{Moshe Schwartz}}
\IEEEauthorblockA{Electrical and Computer Engineering  \\
Ben-Gurion University of the Negev\\
Beer Sheva 8410501, Israel \\
{\it schwartz@ee.bgu.ac.il}}
\and
\IEEEauthorblockN{\textbf{Jehoshua Bruck}}
\IEEEauthorblockA{Electrical Engineering  \\
California Institute of Technology \\
Pasadena, CA 91125, U.S.A. \\
{\it bruck@paradise.caltech.edu}}
}
\maketitle
\begin{abstract}
We study the rate-distortion relationship in the set of permutations
endowed with the Kendall $\tau$-metric and the Chebyshev metric. Our
study is motivated by the application of permutation rate-distortion
to the average-case and worst-case analysis of algorithms for ranking
with incomplete information and approximate sorting algorithms. For
the Kendall $\tau$-metric we provide bounds for small, medium, and
large distortion regimes, while for the Chebyshev metric we present
bounds that are valid for all distortions and are especially accurate
for small distortions. In addition, for the Chebyshev metric, we
provide a construction for covering codes.
\end{abstract}

\section{Introduction}
In the analysis of sorting and ranking algorithms, it is often assumed that complete information is available, that is, the answer to \emph{every} question of the form ``is $x>y$?'' can be found, either by query or computation. A standard and straightforward result in this setting is that, on average, one needs at least $\log_2 n!$ pairwise comparisons to sort a randomly-chosen permutation of length $n$. In practice, however, it is usually the case that only partial information is available. One example is the learning-to-rank problem, where the solutions to pairwise comparisons are learned from data, which may be incomplete, or in big-data settings, where the number of items may be so large as to make it impractical to query every pairwise comparison. It may also be the case that only an approximately-sorted list is required, and thus one does not seek the solutions to all pairwise comparisons. In such cases, the question that arises is what is the quality of a ranking obtained from incomplete data, or an approximately-sorted list.

One approach to quantify the quality of an algorithm that ranks with incomplete data is to find the relationship between the number of comparisons and the average, or worst-case, quality of the output rankings, as measured via a metric on the space of permutations. To explain, consider a deterministic algorithm for ranking $n$ items that makes $nR$ queries and outputs a ranking of length $n$.  Suppose that the true ranking is $\pi$. The information about $\pi$ is available to the algorithm only through the queries it makes. Since the algorithm is deterministic, the output, denoted as $f(\pi)$, is uniquely determined by $\pi$. The ``distortion'' of this output can be measured with a metric $\dist$ as $\dist(\pi,f(\pi))$. The goal is to find the relationship between $R$ and $\dist(\pi,f(\pi))$ when $\pi$ is chosen at random and when it is chosen by an adversary.
 
A general way to quantify the best possible performance by such an algorithm is to use the rate-distortion theory on the space of permutations. In this context, the codebook is the set $\{f(\pi):\pi\in \Sn\}$, where $\Sn$ is the set of permutations of length $n$, and the rate is determined by the number of queries. For a given rate, no algorithm can have smaller distortion than what is dictated by rate-distortion. 

With this motivation, we study rate distortion in the space of
permutations under the Kendall $\tau$-metric and the Chebyshev
metric. Previous work on this topic includes~\cite{WanMazWor13}, which
studies permutation rate-distortion with respect to the Kendall
$\tau$-metric and the $\ell_1$-metric of inversion vectors,
and~\cite{GieSchSto06} which considers Spearman's footrule.

In this work we study rate distortion in the Kendall $\tau$-metric, which counts the number of pairs that are ranked incorrectly, and the Chebyshev metric, which is the largest error in the rank of any item. Our results on the Kendall $\tau$-metric improve upon those presented in~\cite{WanMazWor13}. In particular, for the small distortion regime, as defined later in the paper, we eliminate the gap between the lower bound and the upper bound given in~~\cite{WanMazWor13}; for the large distortion regime, we provide a stronger lower bound; and for the medium distortion regime, we provide upper and lower bounds with error terms. Our study includes both worst-case and average-case distortions as both measures are frequently used in the analysis of algorithms. We also note that permutation rate-distortion results can also be applied to lossy compression of permutations, e.g., rank-modulation signals \cite{JiaMatSchBru09}. 
Finally, we also present covering codes for the Chebyshev metric, where covering codes for the Kendall $\tau$-metric were already presented in \cite{WanMazWor13}. The codes are the covering analog of the error-correcting codes already presented in \cite{TamSch10,JiaSchBru10,MazBarZem13,BarMaz10}. 

The rest of the paper is organized as follows. In
Section~\ref{secPrelim}, we present preliminaries and
notation. Section~\ref{secNonAsymp} contains non-asymptotic results
valid for both metrics under study. Finally, Section~\ref{secKenTau}
and Section~\ref{secCheb} focus on the Kendall $\tau$-metric and the
Chebyshev metric, respectively.

\section{Preliminaries and Definitions}\label{secPrelim}

For a nonnegative integer $n$, let $[n]$ denote the set $ \{
1,\dots,n \} $, and let $\Sn$ denote the set of permutations of
$[n]$. We denote a permutation $\sigma\in S_n$ as
$\sigma=[\sigma_1,\sigma_2,\dots,\sigma_n]$, where the permutation
sets $\sigma(i)=\sigma_i$. We also denote the identity permutation by
$\id=[1,2,\dots,n]$.

The Kendall $\tau$-distance between two permutations
$\pi,\sigma\in\Sn$ is the number of transpositions of adjacent
elements needed to transform $\pi$ into $\sigma$, and is denoted by
$\dk (\pi,\sigma )$. In contrast, the Chebyshev distance between $\pi$
and $\sigma$ is defined as
\[\dc (\pi,\sigma )=\max_{i\in[n]}|\pi(i)-\sigma(i)|.\] 
Additionally, let $\dist (\pi,\sigma )$ denote a generic distance
measure between $\pi$ and $\sigma$.

Both $\dk$ and $\dc$ are invariant; the former is left-invariant and
the latter is right-invariant \cite{DezHua98}. Hence, the size of the
ball of a given radius in either metric does not depend on its
center. The size of a ball of radius $r$ with respect to $\dk$, $\dc$,
and $\dist$, is given, respectively, by $\bk(r)$, $\bc(r)$, and
$\bb(r)$. The dependence of the size of the ball on $n$ is
implicit.

A code $C$ is a subset $C\subseteq\Sn$. For a code $C$ and a
permutation $\pi\in\Sn $, let
\[\dist(\pi,C)=\min_{\sigma\in C}\dist (\pi,\sigma )\]
be the (minimal) distance between $\pi$ and $C$.

We use $\wcm(D)$ to denote the minimum number of codewords required for a worst-case distortion $D$. That is, $\wcm(D)$ is the size of the smallest code $C$ such that for all $\pi\in\Sn$, we have $\dist(\pi,C)\le D$. Similarly, let $\acm(D)$ denote the minimum number of codewords required for an average distortion $D$ under the uniform distribution on $\Sn $, that is, the size of the smallest code $C$ such that
\[
\frac{1}{n!}\sum_{\pi\in\Sn}\dist(\pi,C)\le D.
\]
Note that $\acm(D)\le\wcm(D)$. In what follows, we assume that the
distortion $D$ is an integer. For worst-case distortion, this
assumption does not lead to a loss of generality as the metrics under
study are integer valued.
 
We also define
\begin{align*}
\wcr(D) & =\frac1n\lg\wcm(D), & \acr(D) & =\frac1n\lg\acm(D),\\
\wca(D) & =\frac{1} n \lg\frac{\wcm(D)}{n!}, & \aca(D) & =\frac{1} n \lg\frac{\acm(D)}{n!},
\end{align*}
where we use $\lg$ as a shorthand for $\log_2$. It is clear that
$\wcr(D)=\wca(D)+{\lg n!}/ n $, and that a similar relationship holds
between $\acr(D)$ and $\aca(D)$. The reason for defining $\wca$ and
$\aca$ is that they sometimes lead to simpler expressions compared to $\wcr$ and
$\acr$. Furthermore, $\wca$ (resp.~$\aca$) can be interpreted as the
difference between the number of bits per symbol required to identify
a codeword in a code of size $\wcm$ (resp.~$\acm$) and the number of
bits per symbol required to identify a permutation in $\Sn$.

Throughout the paper, for $\wcm$, $\acm$, $\wca$, $\aca$, $\wcr$, and $\acr$, subscripts $K$ and $C$ denote that the subscripted quantity corresponds to the Kendall $\tau$-metric and the Chebyshev metric, respectively. Lack of subscripts indicates that the result is valid for both metrics.

In the sequel,
the following inequalities~\cite{CohHonLitLob97} will be useful,
\begin{align}
\frac{2^{nH (p )}}{\sqrt{8np (1-p )}} & \le\binom n {pn}\le\frac{2^{nH (p )}}{\sqrt{2\pi np (1-p )}},\label{eq:bino-approx}\\
\sqrt{2\pi n} (n/e )^ n  & <\ \ n!\ \ <\sqrt{2\pi n} (n/e )^ n e^{1/ (12n )},\label{eq:factor-approx}
\end{align}
where $H (\cdot )$ is the binary entropy function, and $0<p<1$. Furthermore, to denote $\lim_{x\to\infty}\frac{f(x)}{g(x)}=1$, we use \[f(x)\sim g(x) \text{ as } x\to\infty,\] or if the variable $x$ is clear from the context, we simply write $f\sim g$.

\section{Non-asymptotic Bounds}\label{secNonAsymp}

In this section, we derive non-asymptotic bounds, that is, bounds that are valid for all positive integers $n$ and $D$. The results in this section apply to both the Kendall $\tau$-distance and the Chebyshev distance as well as any other invariant distance on permutations. 

The next lemma gives two basic lower bounds for $\wcm(D)$
and $\acm(D)$.
\begin{lemma}
For all $n,D\in\N$, 
\begin{align*}
\wcm(D) & \ge\frac{n!} {\bb(D)}, & \acm(D)>\frac{n!} {\bb(D) ( D +1 )}.
\end{align*}
\end{lemma}
\begin{IEEEproof}
Since the first inequality is well known and its proof is clear, we
only prove the second one. Fix $n$ and $D$. Consider a code
$C\subseteq\Sn$ of size $M$ and suppose the average distortion of this
codes is at most $D$. There are at most $M\bb(D)$ permutations $\pi$
such that $\dist(\pi,C)\le D$ and at least $n!-M\bb(D)$ permutations
$\pi$ such that $\dist(\pi,C)\ge D+1$. Hence, $D> (D+1 ) (1-M\bb(D)/n!
)$. The second inequality then follows.
\end{IEEEproof}

In the next lemma, we use a simple probabilistic argument to give an
upper bound on $\wcm(D)$.
\begin{lemma} \label{lemSimpleGenLB}
For all $n,D\in\N$, 
$\wcm(D)\le \lceil n!\ln n!/\bb(D) \rceil$.
\end{lemma}
\begin{IEEEproof}
Suppose that a sequence of $M$ permutations, $\pi_1,\dots,\pi_M$, is
drawn by choosing each $\pi_i$ i.i.d.~with uniform distribution over
$\Sn$. Denote $C=\mathset{\pi_1,\dots,\pi_M}\subseteq\Sn$. The
probability $P_{f}$ that there exists $\sigma\in\Sn$ with $\dist
(\sigma,C )>D$ is bounded by
\begin{align*}
P_{f} & \le\sum_{\sigma\in\Sn }P (\forall i:\dist (\pi_{i},\sigma )>D )
=n! (1-\bb(D)/n! )^{M} \\
 &<n!e^{-M\bb(D)/n!}=e^{\ln n!-M\bb(D)/n!}.
\end{align*}
Let $M= \lceil n!\ln n!/\bb(D) \rceil $ so that $P_{f}<1$. Hence, a
code of size $M$ exists with worst-case distortion $D$.
\end{IEEEproof}

The following theorem by Stein~\cite{Stein1974}, which can be used to obtain existence results for covering codes (see, e.g., \cite{CohHonLitLob97}), to improve the above upper bound. We use a simplified version of this theorem, which is sufficient for our purpose.
\begin{theorem}
\label{th:covbook}
\cite{Stein1974} Consider a set $X$ and a family $\{A_i\}_{i=1}^N$ of sets that cover $X$. Suppose there are integers $N$ and $Q$ such that, $|X|=N$, $|A_i|\le Q$ for all $i$, and each element of $X$ is in at least $Q$ of the sets $A_i$. Then there is subfamily of $\{A_i\}_{i=1}^N$ containing at most $(N/Q)(1+\ln Q)$ sets that cover $X$.
\end{theorem}

In our context $X$ is $\Sn$, $A_i$ are the balls of radius $D$ centered at each permutation, $N=n!$ and $Q=\bb(D)$. Hence, the theorem
implies that
\[
\wcm(D)\le\frac{n!}{\bb(D)} (1+\ln\bb(D) ).
\]
The following theorem summarizes the results of this section.
\begin{theorem}\label{thmNonAsymp}
For all $n,D\in\N$,
\begin{gather}
\frac{n!}{\bb(D)}\le\wcm(D)\le\frac{n!} {\bb(D)}(1+\ln\bb(D) ),\label{eq:thm-worse}\\
\frac{n!}{\bb(D) ( D +1 )}<\acm(D)\le\wcm(D).\label{eq:thm-ave}
\end{gather}

\end{theorem}

\section[The Kendall Tau Metric]{The Kendall $\tau$-Metric}\label{secKenTau}

The goal of this section is to consider the rate-distortion
relationship for the permutation space endowed by the Kendall
$\tau$-metric. First, we find non-asymptotic upper and lower bounds on
the size of the ball in the Kendall $\tau$-metric. Then, in the
following subsections, we consider asymptotic bounds for small,
medium, and large distortion regimes. Throughout this section, we
assume $1\le D<\frac{1}{2}\binom n {2}$ and $n\ge1$. Note that the
case of $D\ge\frac{1}{2}\binom n {2}$ leads to the trivial codes, e.g., $\{
\id, [n,n-1,\dots,1]\} $ and $ \{ \id \} $.

\subsection{Non-asymptotic Results\label{sub:Ken0}}

Let $\mathbb{X}_ n $ be the set of integer vectors
$x=x_{1},x_{2},\dots,x_ n $ of length $n$ such that $0\le x_{i}\le
i-1$ for $i\in[n]$. It is well known (for example, see
\cite{JiaSchBru10}) that there is a bijection between $\mathbb{X}_ n $
and $\Sn $ such that for corresponding elements $x\in\mathbb{X}_ n $
and $\pi\in\Sn $, we have
\[
\dk\left(\pi,\id\right)=\sum_{i=2}^ n x_{i}.
\]
Hence 
\begin{equation}
\bk(r)=\left|\left\{ x\in\mathbb{X}_ n :\sum_{i=2}^ n x_{i}\le r\right\} \right|,\label{eq:SX}
\end{equation}
for $1\le r\le\binom n {2}$. Thus, the number of nonnegative integer
solutions to the equation $\sum_{i=2}^ n x_{i}\le r$ is at least
$\bk(r)$, i.e.,
\begin{equation}
\bk(r)\le\binom{r+n-1}{r}.\label{eq:ken-simple-UB}
\end{equation}

Furthermore, for $\delta\in\Q$, $\delta\geq 0$, such that $\delta n$
is an integer, it can also be shown that
\begin{equation}
\bk\left(\delta n\right)\ge\lfloor1+\delta\rfloor!\lfloor1+\delta\rfloor^{n-\lfloor1+\delta\rfloor}, \label{eq:ken-LB2}
\end{equation}
by noting the fact that the right-hand side of (\ref{eq:ken-LB2})
counts the elements of $\mathbb{X}_ n $ such that
\[
\begin{cases}
0\le x_{i}\le i-1, & \quad\mbox{for }i\le\lfloor1+\delta\rfloor,\\
0\le x_{i}\le\lfloor\delta\rfloor, & \quad\mbox{for }i>\lfloor1+\delta\rfloor,
\end{cases}
\]
and that $\sum_{i\le\lfloor1+\delta\rfloor}\left(i-1\right)+\left(n-\lfloor1+\delta\rfloor\right)\lfloor\delta\rfloor\le\lfloor\delta\rfloor n\le\delta n$.

Next we find a lower bound on $\bk(r)$ with $r<n$. Let $I\left(n,r\right)$
denote the number of permutations in $\Sn $ that are at
distance $r$ from the identity. We have~\cite[p.~51]{Bona2012}
\begin{align*}
I\left(n,r\right) & =\binom{n+r-1}{r}-\left(\binom{n+r-2}{r-1}+\binom{n+r-3}{r-2}\right)\\
 & \quad\ +\sum_{j=2}^{\infty}\left(-1\right)^{j}f_{j},
\end{align*}
where 
\[f_{j}=\binom{n+r-(u_{j}-j)-1}{r-(u_{j}-j)}+\binom{n+r-u_{j}-1}{r-u_{j}},\]
and $u_j=(3j^2+j)/2$.
For $j\ge2$, we have $f_{j}\ge f_{j+1}$. Thus, for $r<n$,
\begin{align*}
I\left(n,r\right) & \ge\binom{n+r-1}{r}\left(1-\frac{r}{n+r-1}\left(1+\frac{r-1}{n+r-2}\right)\right)\\
 & \ge\frac{1}{4}\binom{n+r-1}{r}.
\end{align*}
Hence, for $r<n$, we have
\begin{equation}
\bk(r)\ge\frac{1}{4}\binom{n+r-1}{r}.\label{eq:ken-LB1}
\end{equation}

\begin{figure}
\includegraphics[width=1\columnwidth]{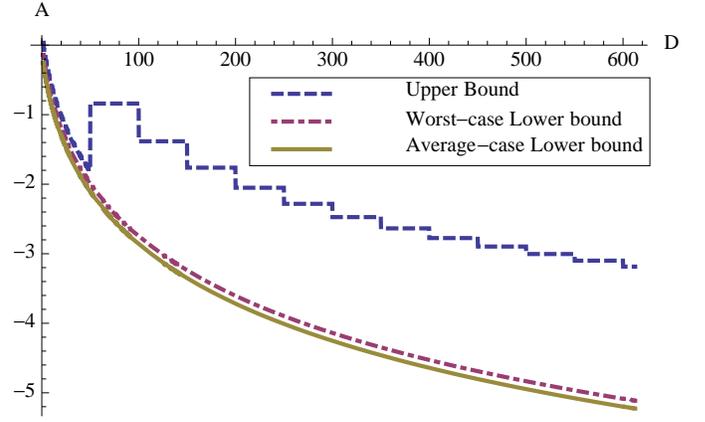}

\caption{Upper bound and lower bounds for $n=50$ from Theorems \ref{thm:LB-A}
and \ref{thm:UB-A}. }
\label{fig:finite-length}
\end{figure}

In the next two theorems, we use the aforementioned bounds on $\bk(r)$
to derive lower and upper bounds on $\wca(D)$ and $\aca(D)$.
\begin{theorem}
\label{thm:LB-A}
For all $n,D\in\N$, and $\delta=D/n$,
\begin{align*}
\wca(D) & \ge-\lg\frac{\left(1+\delta\right)^{1+\delta}}{\delta^{\delta}},\\
\aca(D) & \ge-\lg\frac{\left(1+\delta\right)^{1+\delta}}{\delta^{\delta}}-\frac{\lg n} n.
\end{align*}
\end{theorem}
\begin{IEEEproof}
For the worst-case distortion, we have 
\begin{align*}
\bk(D) & \expl a{\le}\binom{n+\delta n-1}{\delta n} \le\binom{\left(1+\delta\right)n}{\delta n}\\
 & \expl b{\le}\frac{2^{n\left(1+\delta\right)H\left(\frac{1}{1+\delta}\right)}}{\sqrt{2\pi n\delta/\left(1+\delta\right)}} \expl c{\le}2^{n\left(1+\delta\right)H\left(\frac{1}{1+\delta}\right)},
\end{align*}
where $\left(\mathsf{a}\right)$ follows from (\ref{eq:ken-simple-UB}),
$\left(\mathsf{b}\right)$ follows from (\ref{eq:bino-approx}), and
$\left(\mathsf{c}\right)$ follows from the facts that $\delta\ge1/n$
and $n\ge1$. The first result then follows from (\ref{eq:thm-worse}).

For the case of average distortion, we proceed as follows:
\begin{align*}
\bk(D)(D+1)& 
 \le\bk(\delta n)\left(\delta n+1\right)\\
&\le\binom{n+\delta n-1}{\delta n}\left(\delta n+1\right)\\
&=\binom{n+\delta n}{\delta n}\frac{\delta n+1}{1+\delta}\\
 & \expl a{\le}2^{n(1+\delta)H\left(\frac{1}{1+\delta}\right)}\frac{\delta n+1}{\sqrt{2\ppi n\delta(1+\delta)}}\\
 & =2^{n(1+\delta)H\left(\frac{1}{1+\delta}\right)}\sqrt{\frac{2\delta n}\ppi}
 \frac{1+1/(\delta n)}{2\sqrt{n\delta(1+\delta)}}\\
 & \expl b{\le}2^{n\left(1+\delta\right)H\left(\frac{1}{1+\delta}\right)}\sqrt{2\delta n/\ppi},
\end{align*}
where $\left(\mathsf{a}\right)$ follows from (\ref{eq:bino-approx})
and $\left(\mathsf{b}\right)$ is proved as follows. The expression
$\frac{1+1/(\delta n)}{2\sqrt{n\delta(1+\delta)}}$ is decreasing in
$\delta$ for positive $\delta$ and so it is maximized by letting
$\delta=1/n$. Hence,
\[
\frac{1+1/(\delta n)}{2\sqrt{n\delta(1+\delta)}}
\leq
\frac{1}{\sqrt{1+1/n}}\le 1.
\]
Now, using (\ref{eq:thm-ave}) leads to (a stronger version of) the statement in the theorem.\end{IEEEproof}
\begin{theorem}
\label{thm:UB-A}Assume $n,D\in\N$, and let $\delta=D/n$.
We have
\[
\aca(D)\le\wca(D)\le-\lg\frac{\left(1+\delta\right)^{1+\delta}}{\delta^{\delta}}+\frac{3\lg n+12}{2n}.
\]
for $\delta<1$, and 
\[
\aca(D)\le\wca(D)\le-\lg\lfloor1+\delta\rfloor+\frac{1} n \lg\left(ne^{\lfloor1+\delta\rfloor}\ln\lfloor1+\delta\rfloor\right),
\]
for $\delta\ge1$.
\end{theorem}
\begin{IEEEproof}
For $\delta<1$, we have
\begin{align*}
\bk(D) & =\bk\left(\delta n\right)\ge\frac{1}{4}\binom{n+\delta n-1}{\delta n}\\
 & \ge\frac n {4\left(n+\delta n\right)}\binom{n+\delta n}{\delta n}\\
 & \ge\frac{1}{4\left(1+\delta\right)}\cdot\frac{2^{n\left(1+\delta\right)H\left(\frac{1}{1+\delta}\right)}}{\sqrt{8n\delta/\left(1+\delta\right)}}\\
 & =\frac{1}{4}\cdot\frac{2^{n\left(1+\delta\right)H\left(\frac{1}{1+\delta}\right)}}{\sqrt{8n\delta(1+\delta)}}\ge\frac{2^{n\left(1+\delta\right)H\left(\frac{1}{1+\delta}\right)}}{16\sqrt n },
\end{align*}
where the first inequality follows from (\ref{eq:ken-LB1}) and the
last step follows from the fact that $\delta\le1$, and so
$\delta(1+\delta)\le 2$.

Since $\frac{1+\ln x}{x}$ is a decreasing function for $x\ge1$,
we can substitute the lower bound on $\bk(D)$ in (\ref{eq:thm-worse})
to obtain 
\begin{align*}
\wcm(D) & \le\frac{n!16\sqrt{n}}{2^{n\left(1+\delta\right)H\left(\frac{1}{1+\delta}\right)}}\ln\left(\frac{e2^{n\left(1+\delta\right)H\left(\frac{1}{1+\delta}\right)}}{16\sqrt n }\right)\\
 & \stackrel{\mathsf{(a)}}{\le}\frac{n!16 n^{3/2}}{2^{n\left(1+\delta\right)H\left(\frac{1}{1+\delta}\right)}}\left(1+\delta\right)H\left(\frac{1}{1+\delta}\right)\ln2\\
 & \stackrel{\mathsf{(b)}}{\le}\frac{n! 64 n^{3/2}}{2^{n\left(1+\delta\right)H\left(\frac{1}{1+\delta}\right)}},
\end{align*}
where $\mathsf{(a)}$ follows from the fact $e\le 16\sqrt n $ and
$\mathsf{(b)}$ from the fact that for $\delta\le1$, we have $\left(1+\delta\right)H\left(\frac{1}{1+\delta}\right)\ln2\le2\ln2\le4$.
Thus 
\begin{align*}
\wca(D) & \le-\lg\frac{\left(1+\delta\right)^{1+\delta}}{\delta^{\delta}}+\frac{3\lg n+12}{2n}.
\end{align*}

For $\delta\ge1$, by \eqref{eq:ken-LB2} and \eqref{eq:factor-approx}
we have
\begin{align*}
\bk(D) & =\bk(\delta n) \ge\lfloor1+\delta\rfloor!\lfloor1+\delta\rfloor^{n-\lfloor1+\delta\rfloor}\ge\frac{\lfloor1+\delta\rfloor^ n }{e^{\lfloor1+\delta\rfloor}},
\end{align*}
implying
\begin{align*}
\wca(D) & \le\frac{1} n \lg\frac{1+\ln\bk(\delta n)}{\bk(\delta n)}\\
 & \le\frac{1} n \lg\frac{e^{\lfloor1+\delta\rfloor}}{\lfloor1+\delta\rfloor^ n }+\frac{1} n \lg\left(1+n\ln\lfloor1+\delta\rfloor-\lfloor1+\delta\rfloor\right)\\
 & \le\frac{1} n \lg\frac{e^{\lfloor1+\delta\rfloor}}{\lfloor1+\delta\rfloor^ n }+\frac{1} n \lg\left(n\ln\lfloor1+\delta\rfloor\right)\\
 & \le-\lg\lfloor1+\delta\rfloor+\frac{1} n \lg\left(ne^{\lfloor1+\delta\rfloor}\ln\lfloor1+\delta\rfloor\right).
\end{align*}

\end{IEEEproof}
The plots for the expressions given in Theorems \ref{thm:LB-A} and
\ref{thm:UB-A} are given in Figure \ref{fig:finite-length}.

\subsection{Small Distortion}

In this subsection, we consider small distortion, that is, $D=O\left(n\right)$. 

First, suppose $D<n$, or equivalently, $\delta=D/n<1$. The next lemma follows from Lemmas~\ref{thm:LB-A} and \ref{thm:UB-A}.
\begin{lemma}
For $\delta=D/n<1$, we have that
\begin{align}
\wca(D) & =-\lg\frac{\left(1+\delta\right)^{1+\delta}}{\delta^{\delta}}+O\left(\frac{\lg n} n \right),\label{eq:small-distortion1-wc}
\end{align}
and that $\aca(D)$ satisfies the same equation.
\end{lemma}

Next, let us consider the case of $D=\Theta\left(n\right)$. From
(\ref{eq:SX}), it follows that
\[
\bk(k) =\left[z^{k}\right]\frac{1}{1-z}\prod_{i=2}^ n \frac{1-z^{i}}{1-z}=\left[z^{k}\right]\frac{\prod_{i=2}^ n \left(1-z^{i}\right)}{\left(1-z\right)^ n }.
\]
Let 
\begin{align}
g\left(k,n\right) & =\binom{n+k-1}{k}^{-1}\bk(k),\nonumber \\
\gamma\left(z,n\right) & =\sum_{i=0}^{\infty}\Gamma_{i}\left(n\right)z^{i}=\prod_{i=2}^ n \left(1-z{}^{i}\right),\label{eq:gamma}
\end{align}
and 
\[
f\left(z,n\right)=\sum_{i=0}^{\infty}F_{i}\left(n\right)z^{i}=\frac{1}{\left(1-z\right)^ n },
\]
where 
\[
F_{i}\left(n\right)=\binom{n+i-1}{i},
\]
so that 
\[
g\left(k,n\right)=\frac{1}{F_{k}\left(n\right)}\left[z^{k}\right]f\left(z,n\right)\gamma\left(z,n\right).
\]

We use the following theorem to find the asymptotics of $g\left(k,n\right)$
and $\bk(k)$ using the asymptotics of $\gamma\left(z,n\right)$
in Theorem~\ref{thm:asymp2}.

\begin{theorem}
\cite[Theorem 3.1]{MilCom04} \label{thm:asymp1}Let
$f\left(z,n\right)$ and $\gamma\left(z,n\right)$ be two functions
with Taylor series for all $n$,
\begin{align*}
f\left(z,n\right) & =\sum_{i=0}^{\infty}F_{i}\left(n\right)z^{i},\quad\gamma\left(z,n\right)=\sum_{i=0}^{\infty}\Gamma_{i}\left(n\right)z^{i},
\end{align*}
where $F_{i}\left(n\right)>0$ for all sufficiently large $n$. Suppose
\[
g\left(k,n\right)=\frac{1}{F_{k}\left(n\right)}\left[z^{k}\right]f\left(z,n\right)\gamma\left(z,n\right),
\]
and let $n=n\left(k\right)$ be a function of $k$ such that the limit
$\rho=\lim_{k\to\infty}\frac{F_{k-1}\left(n\left(k\right)\right)}{F_{k}\left(n\left(k\right)\right)}$
exists. We have
\[
g\left(k,n\left(k\right)\right)\sim\gamma\left(\rho,n\left(k\right)\right)\mbox{ as }k\to\infty,
\]
provided that
\begin{enumerate}
\item for all sufficiently large $k$ and for all $i$,
\[
\left|\frac{\Gamma_{i}\left(n\left(k\right)\right)}{\gamma\left(\rho,n\left(k\right)\right)}\right|\le p_{i},
\]
where $\sum_{i=0}^{\infty}p_{i}\rho^{i}<\infty$, and
\item there exists a constant $c$, such that for all sufficiently large
$i\le k$ and large $k$,
\[
\left|\frac{F_{k-i}\left(n\left(k\right)\right)}{F_{k}\left(n\left(k\right)\right)}\right|\le c\rho{}^{i}.
\]

\end{enumerate}
\end{theorem}

\begin{theorem}
\label{thm:asymp2}Let $n=n\left(k\right)=\frac{k}{c}+O\left(1\right)$
for a constant $c>0$. Then
\begin{equation}
\bk(k)\sim K_{c}\binom{n+k-1}{k}\label{eq:ball-est}
\end{equation}
as $k,n\to\infty$, where $K_{c}$ is a positive constant equal to
$\lim_{n\to\infty}\gamma\left(c/\left(1+c\right),n\right)$. \end{theorem}
\begin{IEEEproof}
To prove the theorem, we use Theorem~\ref{thm:asymp1}. To do this,
we first let 
\[
\rho=\lim_{k\to\infty}\frac{\binom{n\left(k\right)+k-2}{k-1}}{\binom{n\left(k\right)+k-1}{k}}=\lim_{k\to\infty}\frac{k}{n\left(k\right)+k-1}=\frac{c}{1+c}.
\]
We now turn our attention to Condition 1 of Theorem~\ref{thm:asymp1}.
First, we show that $\gamma\left(\rho,n\left(k\right)\right)$ is
bounded away from 0. We have 
\begin{align*}
\ln\gamma\left(\rho,n\left(k\right)\right) & \ge\sum_{i=2}^{\infty}\ln\left(1-\rho^{i}\right)\ge-\sum_{i=2}^{\infty}\frac{\rho^{i}}{1-\rho^{i}}\\
 & \ge-\sum_{i=2}^{\infty}\frac{\rho^{i}}{1-\rho}=-\frac{\rho^{2}}{\left(1-\rho\right)^{2}},
\end{align*}
where the second inequality follows from the fact that 
\[
\ln\left(1-x\right)=-\sum_{i=1}^{\infty}\frac{x^{i}}{i}\ge-\sum_{i=1}^{\infty}x^{i}=\frac{x}{1-x},
\]
for $0<x<1$. Hence, 
\[
\gamma\left(\rho,n\left(k\right)\right)\ge e^{-\left(\frac{\rho}{1-\rho}\right)^{2}}>0.
\]
To satisfy Condition 1 of Theorem~\ref{thm:asymp1}, it thus suffices
to find $p_{i}'$ such that $\left|\Gamma_{i}\left(n\left(k\right)\right)\right|\le p_{i}'$
and $\sum_{i=0}^{\infty}p_{i}'\rho^{i}<\infty$. For all positive
integers $m$, we have
\begin{align*}
\left|\Gamma_{i}\left(m\right)\right| & = & \biggl|\left[z^{i}\right]\prod_{j=2}^{m}\left(1-z^{j}\right)\biggl| & \le\biggl|\left[z^{i}\right]\prod_{j=2}^{m}\left(1+z^{j}\right)\biggr|\\
 & \le & \biggl|\left[z^{i}\right]\prod_{j=1}^{\infty}\left(1+z^{j}\right)\biggl| & <e^{\pi\sqrt{2/3}\sqrt{i}},
\end{align*}
where the last inequality follows from the facts that $\prod_{j=1}^{\infty}\left(1+z^{j}\right)$
is the generating function for the number of partitions of a positive
integer into distinct parts and that the number of partitions of a
positive integer $i$ is bounded by $e^{\pi\sqrt{2/3}\sqrt{i}}$~\cite[p.~316]{Apo76}. 

We let $p'_{i}=e^{\pi\sqrt{2/3}\sqrt{i}}$ and apply the ratio test
to the sum $\sum_{i=0}^{\infty}p_{i}'\rho^{i}$ to prove its convergence.
Since
\[
\lim_{i\to\infty}\left(p_{i}'\rho^{i}\right)^{1/i}=\lim_{i\to\infty}e^{\pi\sqrt{2/3}/\sqrt{i}}\rho<1,
\]
the sum converges and Condition 1 of Theorem~\ref{thm:asymp1} is
satisfied. Hence, 
\[
\frac{\bk(k)}{\binom{n+k-1}{k}}\sim\gamma\left(\frac{c}{1+c},n\right).
\]

To complete the proof, we show that the limit
$\lim_{n\to\infty}\gamma\left(c/\left(1+c\right),n\right)$ exists and
is positive. This is evident as
$\gamma\left(c/\left(1+c\right),n\right)$ is decreasing and, as shown
before, bounded away from 0.
\end{IEEEproof}

For $D=cn+O\left(1\right)$, we have
\begin{align*}
\frac{1} n \lg\bk(D) & =\frac{1} n \lg\binom{n+D-1}{D}+O\left(\frac{1} n \right)\\
 & =\frac{n+cn+O\left(1\right)} n H\left(\frac{c}{1+c}+O\left(\frac{1} n \right)\right)\\
 & \qquad+O\left(\frac{\lg n} n \right)\\
 & =\left(1+c\right)H\left(\frac{c}{1+c}\right)+O\left(\frac{1} n \right)+O\left(\frac{\lg n} n \right)\\
 & =\left(1+c\right)H\left(\frac{c}{1+c}\right)+O\left(\frac{\lg n} n \right),
\end{align*}
where we used (\ref{eq:ball-est}) for the first step. Using
(\ref{eq:thm-worse}), for $D=cn+O\left(1\right)$, we find
\begin{align*}
\wca(D) & \ge-\frac{1} n \lg\bk(D) =\left(1+c\right)H\left(\frac{c}{1+c}\right)+O\left(\frac{\lg n} n \right)
\end{align*}
and 
\begin{align*}
\wca(D) & \le-\frac{1} n \lg\bk(D)+\frac{1} n \lg\left(1+\ln\bk(D)\right)\\
 & =-\left(1+c\right)H\left(\frac{c}{1+c}\right)+O\left(\frac{\lg n} n \right)
\end{align*}
The derivation for $\aca(cn+O(1))$ is similar. We thus have the following lemma.
\begin{lemma}
For a constant $c>0$ and $D=cn+O(1)$, we have
\begin{equation}
\wca\left(cn+O\left(1\right)\right)=-\lg\frac{\left(1+c\right)^{1+c}}{c^{c}}+O\left(\frac{\lg n} n \right),\label{eq:small-distortion2-wc}
\end{equation}
Furthermore, $\aca(cn+O(1))$ satisfies the same equation.
\end{lemma}

The results given in (\ref{eq:small-distortion1-wc}) and (\ref{eq:small-distortion2-wc})
are given as lower bounds in \cite[Equation~(14)]{WanMazWor13}. We
have thus shown that these lower bounds in fact match the quantity
under study. Furthermore, we have shown that $\aca(D)$
satisfies the same relations.

\subsection{Medium Distortion}

We next consider the medium distortion regime, that is, $D=cn^{1+\alpha}+O\left(n\right)$
for constants $c>0$ and $0<\alpha<1$. For this case, from \cite{WanMazWor13},
we have 
\[
\wca(D)\sim-\lg n^{\alpha},
\]
In this subsection, we improve upon this result by providing upper
and lower bound with error terms.
\begin{lemma}
\label{lem:better-bound} For $D=cn^{1+\alpha}+O(n)$,
where $\alpha$ and $c$ are constants such that $0<\alpha<1$ and
$c>0$, we have
\begin{multline*}
-\lg\left(ecn^{\alpha}\right)+O\left(n^{-\alpha}\right)\le\wca(D)\\
\le-\lg\left(cn^{\alpha}\right)+O\left(n^{-\alpha}+n^{\alpha-1}\right)
\end{multline*}
\end{lemma}
\begin{IEEEproof}
Note that from Theorem \ref{thm:LB-A}, we have 
\begin{align*}
\wca(D) & \ge-\lg\frac{(1+\delta)^{1+\delta}}{\delta^{\delta}}
=\lg\frac{1}{1+\delta}+\lg\left(1+\frac{1}{\delta}\right)^{-\delta}\\
&\ge-\lg\left(e\left(1+\delta\right)\right).
\end{align*}
Let $\delta= D /n=cn^{\alpha}+O\left(1\right)$. We find
\begin{align*}
\wca(D) & \ge-\lg\left(e\left(1+\delta\right)\right)=-\lg e-\lg\left(cn^{\alpha}+O\left(1\right)\right)\\
 & =-\lg\left(ecn^{\alpha}\right)+O\left(n^{-\alpha}\right).
\end{align*}
On the other hand, from Theorem \ref{thm:UB-A},
\begin{align*}
\wca(D) & \le-\lg\left(cn^{\alpha}+O\left(1\right)\right)+\frac{1} n \lg e^{O\left(n^{\alpha}\right)}\\
 & =-\lg\left(cn^{\alpha}\right)+O\left(n^{-\alpha}+n^{\alpha-1}\right)
\end{align*}

\end{IEEEproof}

\subsection{Large Distortion}
\begin{figure}
\includegraphics[width=1\columnwidth]{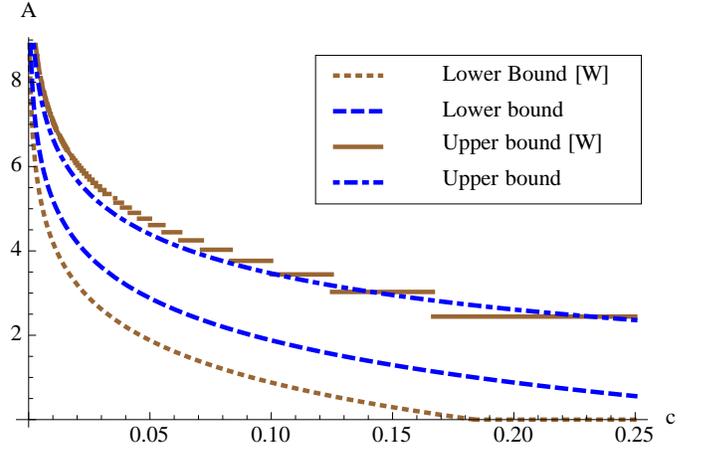}
\caption{Bounds on $\wca(D)+\lg n$ for $D=cn^2+O(n)$ where the error terms are ignored. The bounds denoted by [W] are those from \cite{WanMazWor13}.}
\label{fig:large-dist}

\end{figure}

In the large distortion regime, we have $D=cn^{2}+O(n)$ and $\delta=cn+O\left(1\right)$.
\begin{lemma}
Suppose $D=cn^{2}+O(n)$ for a constant $0<c<\frac{1}{2}$.
We have
\begin{multline*}
-\lg\left(ecn\right)+O\left(\frac{1} n \right)\le\wca(D)\le\\
-\lg\left(ecn\right)+\left(1+c\right)\lg e+O\left(\frac{\lg n} n \right).
\end{multline*}
\end{lemma}
\begin{IEEEproof}
Let $\delta=cn+O\left(1\right)$. Similar to the proof of the lower bound in Lemma~\ref{lem:better-bound}, we have $\wca(D)\ge-\lg\left(e\left(1+\delta\right)\right)$, and thus
\begin{align*}
\wca(D) & \ge-\lg\left(ecn+O\left(1\right)\right)\ge-\lg\left(ecn\right)+O\left(\frac{1} n \right).
\end{align*}
On the other hand, from Theorem \ref{thm:UB-A},
\[
\wca(D)\le-\lg\left(ecn\right)+\left(1+c\right)\lg e+O\left(\frac{\lg n} n \right).
\]
\end{IEEEproof}
From \cite{WanMazWor13}, we have
\begin{multline}
-\lg\left(ecn\right)-1+O\left(\frac{\lg n} n \right)\le\wca(D)\le\\
-\lg\frac n {e\left\lceil 1/\left(2c\right)\right\rceil }+O\left(\frac{\lg n} n \right).\label{eq:wang-cn2}
\end{multline}
These bounds are compared in Figure \ref{fig:large-dist}, where we added the term $\lg n$ to remove dependence on $n$.

\section{The Chebyshev Metric}\label{secCheb}
We now turn to consider the rate-distortion function for the
permutation space under the Chebyshev metric. We start by stating
lower and upper bounds on the size of the ball in the Chebyshev
metric, and then construct covering codes.

\subsection{Bounds}
For an $n\times n$ matrix $A$, the permanent of $A=(A_{i,j})$ is defined as,
\[\per(A)=\sum_{\pi\in\Sn}\prod_{i=1}^n a_{i,\pi(i)}.\]
It is well known \cite{Klo08,Sch09} that $\bc(r)$ can be expressed as the
permanent of the $n\times n$ binary matrix $A$ for which
\begin{equation}\label{eq:ChebA}
A_{i,j}=\begin{cases}
1 & \abs{i-j}\leq r\\
0 & \text{otherwise.}
\end{cases}
\end{equation}


According to Br\'{e}gman's Theorem (see \cite{Bre73}), for any
$n\times n$ binary matrix $A$ with $r_i$ $1$'s in the $i$-th row
\[\per(A)\leq \prod_{i=1}^n(r_i !)^{\frac{1}{r_i}}.\]
Using this bound we can state the following lemma (partially given in \cite{Klo08} and extended in \cite{TamSch10}).

\begin{lemma}
\label{lemChebBallUB}
\cite{TamSch10}
For all $0\leq r\leq n-1$,
\[
\bc(r)\leq \begin{cases}
\parenv{(2r+1)!}^{\frac{n-2r}{2r+1}}
\prod_{i=r+1}^{2r}(i!)^{\frac{2}{i}}, & 0\leq r\leq \frac{n-1}{2}, \\
\parenv{n!}^{\frac{2r+2-n} n }
\prod_{i=r+1}^{n-1}(i!)^{\frac{2}{i}}, & \frac{n-1}{2} \leq r \leq n-1.
\end{cases}
\]
\end{lemma}

The following lower bound was given in \cite{Klo08}.
\begin{lemma}
\cite{Klo08}
For all $0\le r\le \frac{n-1}{2}$, 
\[
\bc(r)\ge
\frac{\left(2r+1\right)^ n }{2^{2r}}\frac{n!}{n^ n }.
\]
\end{lemma}

We extend this lemma to the full range of parameters.
\begin{lemma}\label{lemChebBallLB}
For all $0\le r\le n-1$, 
\[
\bc(r)\ge\begin{cases}
\frac{\left(2r+1\right)^ n }{2^{2r}}\frac{n!}{n^ n }, & 0\leq r\leq \frac{n-1}{2}, \\
\frac{n! }{2^{2(n-r)}}, & \frac{n-1}{2} \leq r \leq n-1.
\end{cases}
\]
\end{lemma}
\begin{IEEEproof}
Only the second claim requires proof, so suppose that $(n-1)/2\le r\leq
n-1$. The proof follows the same lines as the one appearing in
\cite{Klo08}. Let $A$ be defined as in \eqref{eq:ChebA}, and let $B$
be an $n\times n$ matrix with
\[
B_{i,j}=\begin{cases}
2, & i+j\le n-r,\\
2, & i+j\ge n+r+2,\\
A_{i,j}, & \text{otherwise.}
\end{cases}
\]
We observe that $B/n$ is doubly stochastic. It follows that
\begin{align*}
\bc(r)& =\per(A)\ge\frac{\per(B)}{2^{2(n-r)}}
 \ge\frac{n^ n }{2^{2(n-r)}}\per\left(\frac{B}n\right)\\
 & \ge\frac{n! }{2^{2(n-r)}},
\end{align*}
where the last inequality follows from Van der Waerden's Theorem \cite{Min78}.
\end{IEEEproof}


\begin{theorem}\label{thmChebBounds}
Let $n\in\N$, and let $0<\delta < 1$ be a constant rational number such that
$\delta n$ is an integer. Then
\[ \wcr_C(D)\geq \begin{cases}
\lg \frac{1}{2\delta}+2\delta\lg \frac e 2+O(\lg n/n), & 0 < \delta \leq \frac{1}{2}\\
2\delta\lg \delta+2(1-\delta)\lg  e+O(\lg n/n), & \frac{1}{2}\leq \delta \leq 1
\end{cases}
\]
and
\[ \wcr_C(D)\leq \begin{cases}
\lg\frac{1}{2\delta}+2\delta+O(\lg n/n), & 0 < \delta \leq \frac{1}{2}\\
2(1-\delta)+O(\lg n/n), & \frac{1}{2}\leq \delta \leq 1
\end{cases}
\]
Furthermore, the same bounds also hold for $\acr_C(D)$.
\end{theorem}
\begin{IEEEproof}
First, we prove the lower bound for $\wcr_C(D)$ using Theorem~\ref{thmNonAsymp}, which
states $\wcm_C(D) \ge n!/\bc(D)$, and Lemma~\ref{lemChebBallUB}. Let
\begin{align*}
T_{1} & =\left((2D+1)!\right)^{(n-2D)/(2D+1)},\\
T_{2} & =\prod_{i=D+1}^{2D}(i!)^{2/i}.
\end{align*}
We have
\begin{align*}
\lg T_{1} & =\frac{n-2\delta n}{2\delta n+1}\lg(2\delta n+1)!\\
 & =\frac{n-2\delta n}{2\delta n+1}\left((2\delta n+1)\lg\left(\frac{2\delta n+1}{e}\right)+O(\lg n)\right)\\
 & =\left(n-2\delta n\right)\lg\left(\frac{2\delta n+1}{e}\right)+O(\lg n)\\
 & =(n-2\delta n)\lg(2\delta n/e)+O(\lg n),
\end{align*}
and
\begin{align*}
\lg T_{2} & =2\sum_{i=\delta n+1}^{2\delta n}\frac{1}{i}\lg i!=2\sum_{i=\delta n+1}^{2\delta n}\left(\lg\frac{i}{e}+O\left(\frac{\lg i}{i}\right)\right)\\
 & =2\sum_{i=\delta n+1}^{2\delta n}\lg i-2\delta n\lg e+O(\lg n)\\
 & =2\lg\frac{(2\delta n)!}{(\delta n)!}-2\delta n\lg e+O(\lg n)\\
 & =2\delta n+2\delta n\lg(2\delta n/e)-2\delta n\lg e+O(\lg n).
\end{align*}
From these expressions and Lemma~\ref{lemChebBallUB}, it follows that
\[\frac{1}{n}\lg\bc(D)\le\lg(2\delta n/e)+2\delta\lg(2/e)+O(\lg n/n).\]
The lower bound for $0<\delta\le1/2$ then follows from
Theorem~\ref{thmNonAsymp}. The proof of the lower bound for
$1/2<\delta\le 1$ is similar.

Next, we prove the upper bound for $\wcr_C(D)$. From Lemma~\ref{lemSimpleGenLB}, we have
\begin{align*}
\wcm_C (D) & \le\frac{n!\ln n!}{\bc(D)}.
\end{align*}
Hence, for $0\le D\le (n-1)/2$,
\begin{align}
\wcr_C (D) & \le\frac{1} n \lg\left(\frac{n!\ln n!}{\bc(\delta n)}\right)\nonumber\\
 & \le\frac{1} n \lg\left(\frac{2^{2\delta n}n^ n }{\left(2\delta n+1\right)^ n }\right)+O\left(\frac{\lg n} n\right) \nonumber\\
& \le\lg\frac{1}{2\delta}+2\delta++O\left(\frac{\lg n} n\right)\label{eq:cheb-UB-1}
\end{align}
where we have used Lemma~\ref{lemChebBallLB} for the second inequality.

Similarly, for $(n-1)/2< D\le n$,
\begin{align}
\wcr_C (D) & \le\frac{1} n \lg 2^{2n(1-\delta)}+O\left(\frac{\lg n} n\right) \nonumber\\
& \le2(1-\delta)+O\left(\frac{\lg n} n\right) .\label{eq:cheb-UB-2}
\end{align} 
The proof of the lower bound for $\acr_C (D)$  is similar to that of $\wcr_C (D)$ except that we use $\acm(D)>n!/(\bb(D)(D+1))$ from Theorem~\ref{thmNonAsymp}. The proof of the upper bound for $\acr_C (D)$ follows from the fact that $\acr_C (D)\le \wcr_C (D)$.
\end{IEEEproof}

\begin{figure}
\includegraphics[width=1\columnwidth]{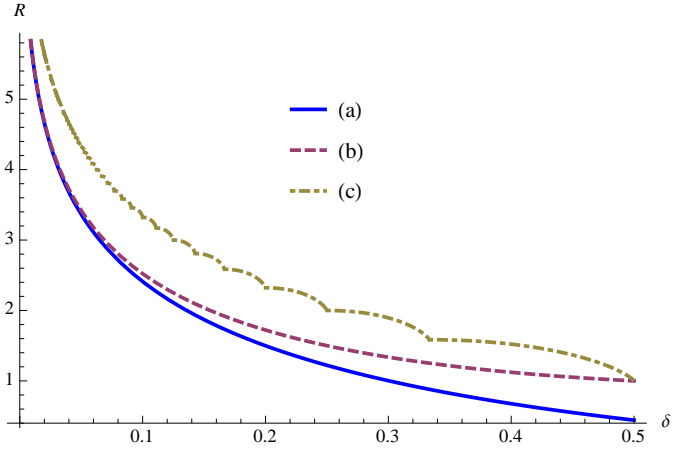}

\caption{Rate-distortion in the Chebyshev metric: The lower and upper bounds of Theorem~\ref{thmChebBounds}, (a) and (b), and the rate of the code construction, given in Theorem~\ref{thmConstruction}, (c).}
\label{fig:chebyshev}

\end{figure}

\subsection{Code Construction}
Let $A=\mathset{a_1,a_2,\dots,a_m}\subseteq [n]$ be a subset of
indices, $a_1<a_2<\dots<a_m$. For any permutation $\sigma\in \Sn$ we
define $\sigma|_A$ to be the permutation in $\Sm$ that preserves the
relative order of the sequence $\sigma(a_1),\sigma(a_2),
\dots,\sigma(a_m)$. Intuitively, to compute $\sigma|_A$ we keep only
the \emph{coordinates} of $\sigma$ from $A$, and then relabel the
entries to $[m]$ while keeping relative order. In a similar fashion we
define
\[\sigma|^A= \parenv{\sigma^{-1}|_A}^{-1}.\]
Intuitively, to calculate $\sigma|^A$ we keep only the \emph{values} of $\sigma$ from $A$, and then relabel the entries to $[m]$ while keeping relative order.

\begin{example}
Let $n=6$ and consider the permutation
\[\sigma=[6,1,3,5,2,4].\]
We take $A=\mathset{3,5,6}$. We then have
\[\sigma|_A=[2,1,3],\]
since we keep positions $3$, $5$, and $6$, of $\sigma$, giving us
$[3,2,4]$, and then relabel these to get $[2,1,3]$.

Similarly, we have
\[\sigma|^A=[3,1,2],\]
since we keep the values $3$, $5$, and $6$, of $\sigma$, giving us
$[6,3,5]$, and then relabel these to get $[3,1,2]$.
\end{example}

\begin{construction}
\label{con:con1}
Let $n$ and $d$ be positive integers, $1\leq d\leq n-1$. Furthermore,
we define the sets
\[A_i=\mathset{i(d+1)+j ~|~ 1\leq j\leq d+1 }\cap [n],\]
for all $0\leq i \leq \floorenv{(n-1)/(d+1)}$.
We now construct the code $C$ defined by
\[C=\mathset{\sigma\in \Sn ~\left|~ \text{$\sigma|^{A_i}=\id$ for all $i$}\right.}.\]
\end{construction}

We note that this construction may be seen as a dual of the
construction given in \cite{WanMazWor13}.

\begin{theorem}
Let $n$ and $d$ be positive integers, $1\leq d \leq n-1$. Then the code
$C\subseteq \Sn$ of Construction \ref{con:con1} has covering radius
exactly $d$ and size
\[M=
\frac{n!}{(d+1)!^{\floorenv{n/(d+1)}}(n\bmod (d+1))!}.\]
\end{theorem}
\begin{IEEEproof}
Let $\sigma\in \Sn$ be any permutation. We let $I_i$ denote the
indices in which the elements of $A_i$ appear in $\sigma$. Let us now
construct a new permutation $\sigma'$ in which the elements of $A_i$
appear in indices $I_i$, but they sorted in ascending order. Thus
\[\sigma'|^{A_i}=\id,\]
for all $i$, and so $\sigma'$ is a codeword in $C$.

We observe that if $\sigma(j)\in A_i$, then $\sigma'(j)\in A_i$ as well. It follows that
\[\abs{\sigma(j)-\sigma'(j)}\leq d\]
and so 
\[\dc(\sigma,\sigma')\leq d.\]

Finally, we contend the permutation $\sigma=[n,n-1,\dots,1]$ is at
distance exactly $d$ from the code $C$. We note we already know that
there is a codeword $\sigma'\in C$ such that $\dc(\sigma,\sigma')\leq
d$. We now show there is no closer codeword in $C$. Let us attempt to build
such a permutation $\sigma''$. Consider $\sigma(n)=1$. The value of
$\sigma''(n)$ is in $A_i$ for some $i$, and since $\sigma''$ is a codeword,
$\sigma''(n)$ must be the largest in $A_i$. Thus
\[\sigma''(n)\in \mathset{ \max(A_i) ~|~ 1\leq i\leq \ceilenv{n/(d+1)}}
\geq d+1.\]
It follows that
\[\abs{\sigma''(n)-\sigma(n)}=d\]
and so
\[\dc(\sigma,\sigma'')\geq d.\]
\end{IEEEproof}
The code construction has the following asymptotic form:
\begin{theorem}\label{thmConstruction}
The code from Construction \ref{con:con1} has the following asymptotic
rate,
\[R=H\parenv{\delta\floorenv{\frac{1}{\delta}}}+\delta\floorenv{\frac{1}{\delta}}\lg \floorenv{\frac{1}{\delta}},\]
where $H$ is the binary entropy function. 
\end{theorem}

The bounds given in Theorem~\ref{thmChebBounds} and the rate of the code construction, given in Theorem~\ref{thmConstruction}, are shown in Figure~\ref{fig:chebyshev}.

\bibliographystyle{IEEEtranS}
\bibliography{allbib}



  


\end{document}